\documentclass[manuscript]{aastex}
\usepackage{natbib}
\usepackage{color}

\newcommand{\ensuretext[1]}{\ensuremath{\text{#1}}}

\newcommand{\g}{\ensuremath{\gamma}}%
\newcommand{\fermi}{\textsl{Fermi}/LAT}
\newcommand{\hess}{\textsc{H.E.S.S.}}
\newcommand{\hessjlong}{HESS\,J1641$-$463}
\newcommand{\hessj}{J1641$-$463}
\newcommand{\fourty}{HESS\,J1640$-$465}

\newcommand{\fluxHESSaatOneTeV}{\ensuremath{\phi_0 =  (3.91 \pm 0.69_{\rm stat} \pm 0.78_{\rm sys}) \times 10^{-13} \rm cm^{-2}s^{-1}TeV^{-1}}}
\newcommand{\intfluxHESSaoneTeV}{\ensuremath{\phi\,(\rm E > 1\, TeV) = (3.64 \pm 0.44_{\rm stat} \pm 0.73_{\rm sys}) \times 10^{-13} \rm cm^{-2}s^{-1}}}

\newcommand{\GammaHESSa}{\ensuremath{\Gamma = 2.07 \pm 0.11_{\rm stat} \pm 0.20_{\rm sys}}}

\newcommand{\RAfitted}{\ensuremath{\rm \alpha_{J2000} = 16^{h} 41 ^{m} 2.1^{s} \pm 3.0 ^{s}_{\rm stat} \pm 1.9 ^{s}_{\rm sys}}}
\newcommand{\DECfitted}{\ensuremath{\rm \delta_{J2000} = -46^{\circ} 18' 13'' \pm 35''_{\rm stat} \pm 20''_{\rm sys}}}

\slugcomment{Draft version}

\shorttitle{Discovery of \hessjlong}
\shortauthors{H.E.S.S. Collab.}

\begin{document}


\title{Discovery of the hard spectrum VHE \g-ray source \hessjlong}


\author{H.E.S.S. Collaboration,
A.~Abramowski\altaffilmark{1},
F.~Aharonian\altaffilmark{2,3,4},
F.~Ait Benkhali\altaffilmark{2},
A.G.~Akhperjanian\altaffilmark{5,4},
E.O.~Ang\"uner\altaffilmark{6},
M.~Backes\altaffilmark{7},
S.~Balenderan\altaffilmark{8},
A.~Balzer\altaffilmark{9},
A.~Barnacka\altaffilmark{10,11}
% ...
}

\altaffiltext{1}{Universit\"at Hamburg, Institut f\"ur Experimentalphysik, Luruper Chaussee 149, D 22761 Hamburg, Germany}
\altaffiltext{2}{Max-Planck-Institut f\"ur Kernphysik, P.O. Box 103980, D 69029 Heidelberg, Germany}
\altaffiltext{3}{Dublin Institute for Advanced Studies, 31 Fitzwilliam Place, Dublin 2, Ireland}
\altaffiltext{4}{National Academy of Sciences of the Republic of Armenia,  Marshall Baghramian Avenue, 24, 0019 Yerevan, Republic of Armenia}
\altaffiltext{5}{Yerevan Physics Institute, 2 Alikhanian Brothers St., 375036 Yerevan, Armenia}
\altaffiltext{6}{Institut f\"ur Physik, Humboldt-Universit\"at zu Berlin, Newtonstr. 15, D 12489 Berlin, Germany}
\altaffiltext{7}{University of Namibia, Department of Physics, Private Bag 13301, Windhoek, Namibia}
\altaffiltext{8}{University of Durham, Department of Physics, South Road, Durham DH1 3LE, U.K.}
\altaffiltext{9}{GRAPPA, Anton Pannekoek Institute for Astronomy, University of Amsterdam,  Science Park 904, 1098 XH Amsterdam, The Netherlands}
\altaffiltext{10}{Obserwatorium Astronomiczne, Uniwersytet Jagiello{\'n}ski, ul. Orla 171, 30-244 Krak{\'o}w, Poland}
\altaffiltext{11}{now at Harvard-Smithsonian Center for Astrophysics,  60 Garden St, MS-20, Cambridge, MA 02138, USA}
% ...

\email{Igor Oya$-$igor.oya.vallejo@desy.de, Sabrina Casanova$-$sabrina.casanova@ifj.edu.pl, sabrina.casanova@mpi-hd.mpg.de}


\begin{abstract}
  This letter reports the discovery of a remarkably hard spectrum
  source, \hessjlong, by the High Energy Stereoscopic System
  (H.E.S.S.) in the very high energy (VHE) domain.
  \hessjlong\ remained unnoticed by the usual analysis techniques due
  to confusion with the bright nearby source \fourty. It emerged at a
  significance level of 8.5 standard deviations after restricting the
  analysis to events with energies above 4 TeV. It shows a moderate
  flux level of \intfluxHESSaoneTeV, corresponding to 1.8\% of the
  Crab Nebula flux above the same energy, and a hard spectrum with a
  photon index of \GammaHESSa. It is a point-like source, although an
  extension up to a Gaussian width of $\sigma = 3$ arcmin cannot be
  discounted due to uncertainties in the H.E.S.S. point-spread function. The VHE \g-ray
  flux of \hessjlong\ is found to be constant over the observed period
  when checking time binnings from the year-by-year to the 28 minute
  exposure timescales. \hessjlong\ is positionally coincident with
  the radio supernova remnant SNR G338.5+0.1. No X-ray candidate
  stands out as a clear association; however, {\it Chandra} and
  XMM-{\it Newton} data reveal some potential weak
  counterparts. Various VHE \g-ray production scenarios are
  discussed. If the emission from \hessjlong\ is produced by cosmic
  ray protons colliding with the ambient gas, then their spectrum must
  extend close to 1 PeV. This object may represent a source
  population contributing significantly to the galactic cosmic ray
  flux around the knee.
\end{abstract}


\section{Introduction}
\label{Introduction}

The large field of view (FoV) of the High Energy Stereoscopic System
(\hess), together with its stereoscopic observation strategy, allowed
the discovery of tens of very high energy (VHE, $\ge$ 0.1 TeV) \g-ray
sources\footnote{See http://tevcat.uchicago.edu/ for an updated list
  of VHE \g-ray sources.}  by scanning a large fraction of the
Galactic plane \citep{2005Sci...307.1938A, 2013arXiv1307.4690C}. With
deeper exposures, more VHE \g-ray sources are detected, although source
confusion begins to be problematic. Complementing the spatial search
for new sources, an investigation into energy bands can provide an
additional powerful tool for new discoveries. In this work, it will be
shown how this method allowed for the detection of a new object,
\hessjlong\ (hereafter, \hessj), previously hidden in the
tails of the much brighter object \fourty. Interestingly, the newly
discovered source exhibits one of the hardest spectra observed in VHE
\g-rays, allowing its detection at higher energies, where the two
sources are clearly separated. Hereafter, the observations and the
analysis technique that led to the discovery of \hessj\ are
described. Finally, a discussion of plausible counterparts of this
source at other wavelengths is presented.

\section{H.E.S.S. observations and results}
\label{discovery}

H.E.S.S. is an array of five imaging atmospheric Cherenkov telescopes
located in the Khomas Highland of Namibia, 1800 m above sea level. In
the initial phase of the H.E.S.S. project, during which the data
described here were taken, the array was composed of four 13 m
diameter telescopes. Extensive air showers are measured with
an average energy resolution of 15\% and an angular
resolution better than 0.1$^{\circ}$ \citep{2006A&A...457..899A} for a
typical energy of 1 TeV. The trigger energy threshold is about 100 GeV
and increases with higher zenith angle \citep{2004APh....22..285F}.

\begin{figure}
  \centering
  \includegraphics[width=6.0in]{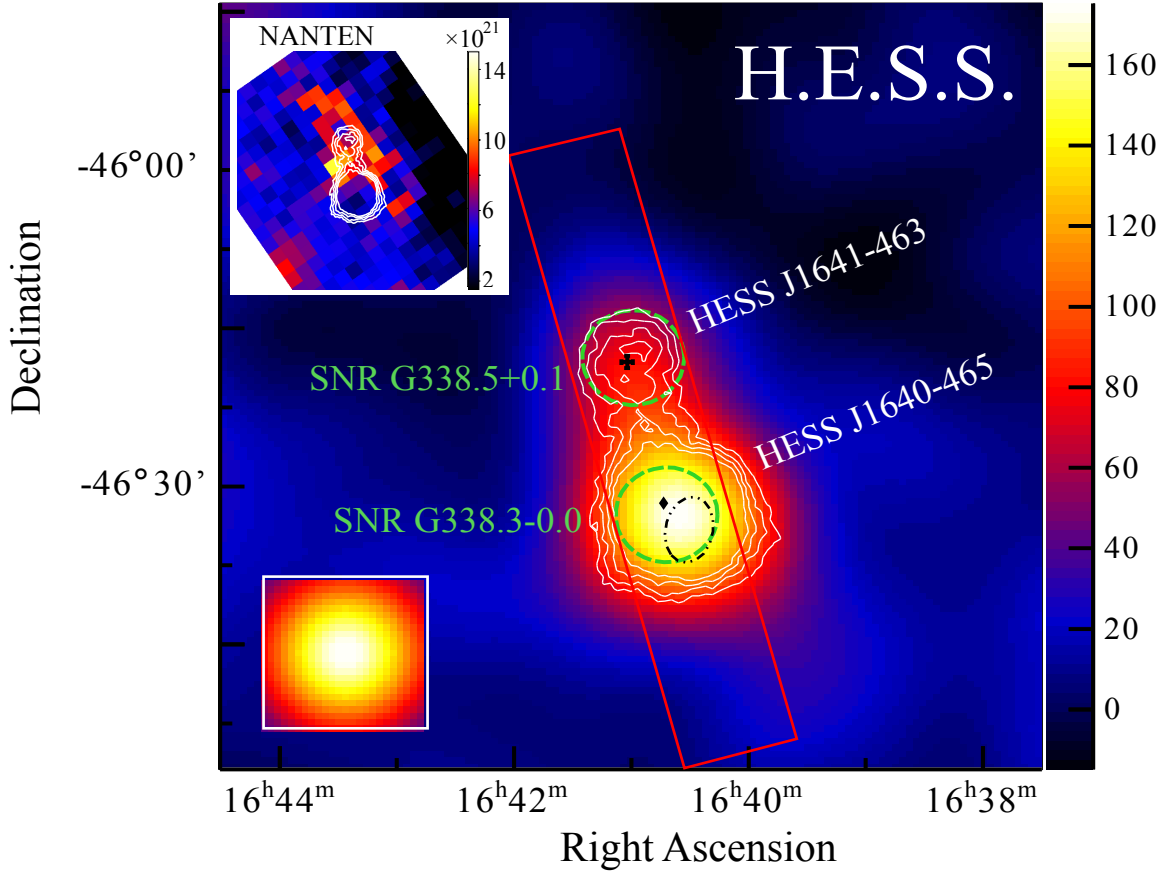}
  \caption{Map of excess events with energies E\,$>$\,4\,TeV for the
    region around \hessj\ smoothed with a Gaussian of width
    0.085$^{\circ}$, corresponding to the 68\% containment radius of
    the instrument PSF. The white contours indicate the significance of
    the emission at the 5, 6, 7, and $8\sigma$ level. The black cross
    indicates the value and uncertainty of the best fit position of
    the source, the green dashed circles show the positions and
    approximate extensions of the two nearby SNRs, the black diamond
    the position of PSR J1640$-$4631, the dash$-$dotted black ellipse the
    95\% confidence error position of 1FHL J1640.5−4634, and the red
    box indicates the area for the extraction of the profiles shown in
    Figure\ \ref{Slices}. The color scale is in units of counts
    per smoothing Gaussian width. The \hess\ PSF is shown inside the
    white box. The upper left inset shows a map of the distribution of
    the column density of molecular hydrogen in units of cm$^{-2}$,
    estimated from the NANTEN CO(1$-$0) data, together with the
    \hess\ significance contours.}
  \label{ExcessMap}
\end{figure}

\hessj\ remained unnoticed by the standard source detection techniques
due to its low brightness and its proximity to the bright
source \fourty\ \citep{2014MNRAS.439.2828A}. During a study of a
possible energy-dependent morphology of \fourty, a collection of
images for events with energies above a set of energy thresholds ($E$
$>$ 1, 2, 3, 4, and 5 TeV) was created. \hessj\ was not visible in the
original images of the \fourty\ FoV, as those images included no energy
cut in the events, and thus were dominated by the much more numerous
low-energy events coming from the brighter \fourty. Thanks to the
improved \hess\ point-spread function (PSF) at higher energies, and to
its hard spectrum, \hessj\ was clearly visible in the highest energy sky
maps, where the contamination from \fourty\ was low. This discovery
triggered further H.E.S.S. observations, allowing the firm
establishment of a new VHE \g-ray source. The VHE \g-ray excess image
obtained for $E$ $>$ 4 TeV is shown in Figure\ \ref{ExcessMap}, where the
background level is estimated following the ring background model
\citep{2007A&A...466.1219B}.

The observations of the FoV around \hessj\ were carried out from 2004
to 2011, corresponding to an acceptance-corrected live time of 72
hr, after quality selection criteria were applied as in
\cite{2006A&A...457..899A}. The data were analyzed with the
methods described in \cite{2006A&A...457..899A}\footnote{The
  H.E.S.S. hap-12-03 analysis software package with version 32
  instrument response tables was used.}. The events were reconstructed
using the Hillas parameter technique \citep{hillas1985}. The
results were cross-checked using two independent analysis methods
\citep{2009APh....31..383O, 2009APh....32..231D}.

The position of \hessj\ (together with the nearby \fourty) was
determined by fitting a two-dimensional double-Gaussian model
convolved with the H.E.S.S. PSF to the two-dimensional ON-source
excess event distribution for $E$ $>$ 4 TeV, energies at which source
confusion with \fourty\ is mitigated. The centroid of the Gaussian
corresponding to the location of \hessj\ was found to be \RAfitted,
\DECfitted. The source is found to be point-like, but a slightly
extended morphology up to a width of $\sigma$ = 3 arcmin cannot
be ruled out due to uncertainties in the H.E.S.S. PSF.

\begin{figure}
  \centering
  \includegraphics[width=4.2in]{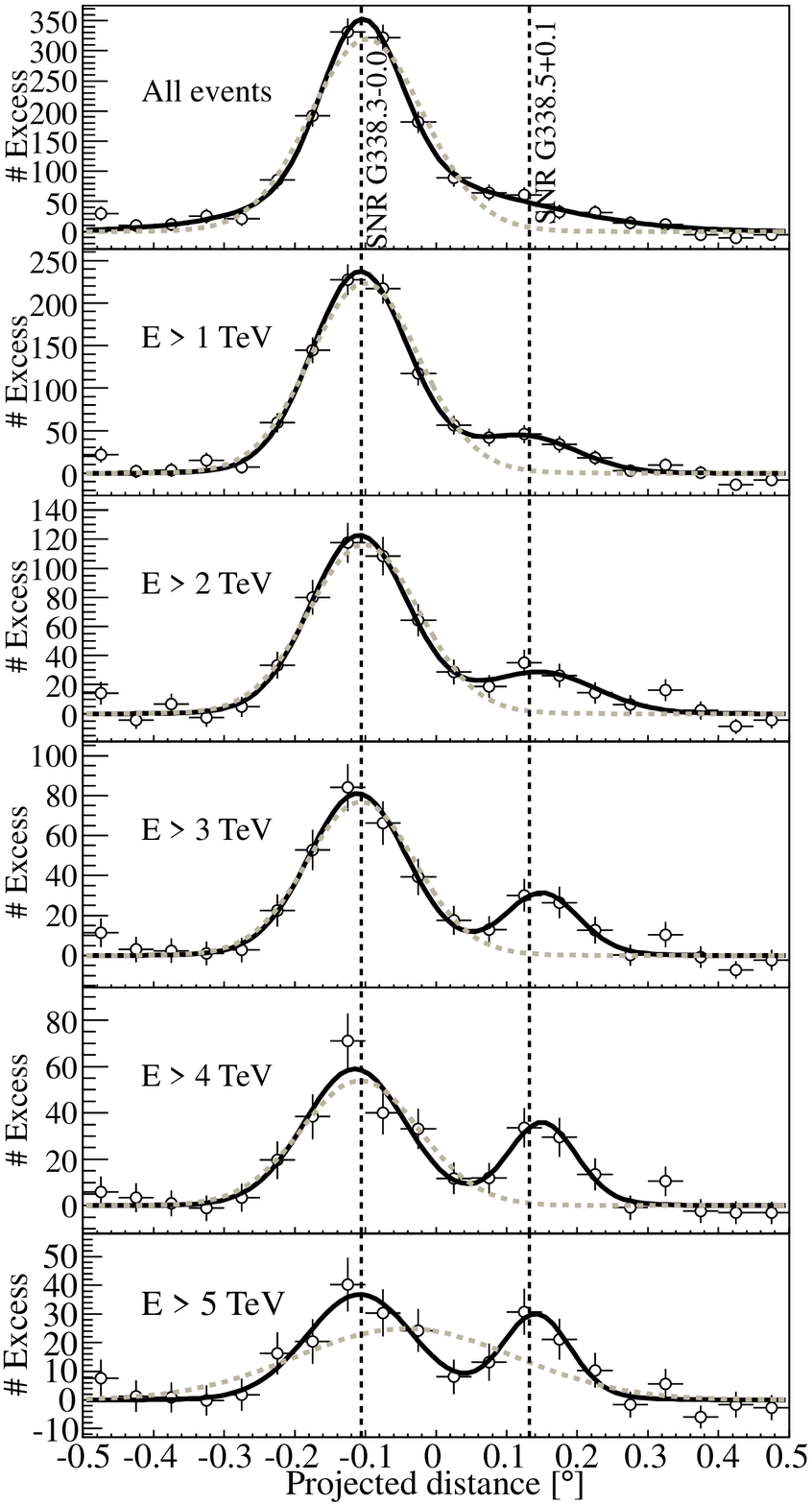}
  \caption{Distribution of VHE $\g$-ray excess profiles and Gaussian
    fits (convolved with the instrument PSF) for the red rectangular
    slice shown in Figure\ \ref{ExcessMap}. Vertical lines show the
    position of the SNR 338.3$-$0.0 and G338.5+0.1. Fits using a
    single and a double Gaussian function are shown in dashed and
    solid lines, respectively. Note that the energy dependence of the
    PSF is taken into account in the fits.}
  \label{Slices}
\end{figure}

Figure~\ref{Slices} shows the projection of the excess
events in the rectangular region shown in Figure \ref{ExcessMap} for
different energy bands. An $F$-test \citep{1971stph.book.....M} was
performed comparing the single Gaussian model fits with the
double-Gaussian fits. For all the energy bands, the null hypothesis can
be rejected at significance levels of $3.6-4.3\sigma$,
thus clearly favoring the double-Gaussian model.

In order to minimize the contamination from \fourty, hard cuts were
used, which imply a cut on $\theta^{2}$ (the square of the angular
difference between the reconstructed shower direction and the source
position) of 0.01 deg$^{2}$, and on the individual image charge in
photo-electrons of 200. The source is detected with a statistical
significance of $8.5\sigma$ above 4 TeV, determined by using Equation
(17) in \cite{1983ApJ...272..317L} after background suppression with
the reflected background model \citep{2007A&A...466.1219B}.

The differential VHE \g-ray spectrum of \hessj, derived using the
forward-folding technique \citep{2001A&A...374..895P}, is compatible
with a power-law function ${\rm d}N/{\rm d}E = \phi_0 \times (E/1
\,\rm TeV)^{-\Gamma}$ with \fluxHESSaatOneTeV\ and \GammaHESSa\ for
the energy range from 0.64 to 100\,TeV. The flux level is
\intfluxHESSaoneTeV, corresponding to 1.8\% of the Crab Nebula flux
above the same energy. At those energies, the estimated total
contamination from \fourty\ is 15\,$\pm$\,6\,\%, reduced at higher
energies (4\,$\pm$\,3\,\% at $E$\,$>$\,4\,TeV). A fit by a power law
with an exponential cutoff is not statistically justified given the low
flux level of \hessj. A fit to a constant value of the
period-by-period\footnote{A \hess\ observing period is the period
  between two full moons.} light curve for energies above 0.64 TeV
yields a $\chi^2$/d.o.f. = 11.7/14, with a $p$-value of 63\%. No
variability can be seen in other time binnings (from
year-by-year to 28 minute exposures).

\section{Search for counterparts at other wavelengths}

\subsection{Radio Observations}\label{radio_sec}

\hessj\ is found within the bounds of SNR\,G338.5+0.1
\citep{2009BASI...37...45G}. This SNR is located at $\alpha_{J2000}$ =
16$^{h}$40$^{m}$59$^{s}$, $\delta_{J2000}$ = $-$46$^{\circ}$17.8$'$,
has a roughly circular morphology, and shows a flux density at 1 GHz
of $\approx12$ Jy \citep{2009BASI...37...45G}. A diameter between 5$'$
(the most obvious non thermal emission region reported in
\citealt{1996A&AS..118..329W}) to 9$'$ \citep{2009BASI...37...45G} for
G338.5+0.1 is assumed in this work, and the latter is displayed in
Figure~\ref{ExcessMap}.  \cite{2007A&A...468..993K} conclude that the
source is located at a distance of 11 kpc,\footnote{Although
  \cite{1970AuJPA..14..133S} report a closer distance of 5.3 kpc, in
  this work it is assumed a distance of 11 kpc as reported by
  \cite{2007A&A...468..993K}.} which implies a physical size between
$\approx16$ to $\approx30$ pc. Assuming that G338.5+0.1 is in the
Sedov$-$Taylor phase, the Sedov solution (see,
e.g., \citealt{2001PhDT........78V}) is used to estimate its age; with
an explosion energy of $10^{51}$ erg and the density of the external
medium between 0.1 and 1 cm$^{-3}$, the age of the SNR would correspond
to 1.1$-$3.5 kyr and 5$-$17 kyr for 16 pc and 30 pc diameter,
respectively.


The distribution of molecular gas around \hessj\ is shown
in the top left inset of Figure~\ref{ExcessMap}.  This distribution is
obtained by integrating the CO 1$\rightarrow$0 rotational line
emission, measured with NANTEN, over a range in velocity between $-$40
km s$^{-1}$ to $-$30 km s$^{-1}$
\citep{2001PASJ...53.1003M,2004ASPC..317...59M}. The choice of this
range is motivated by the presence of dense molecular cloud clumps in
the region, mapped with various NH$_{3}$ emission lines with the MOPRA
survey at those velocities \citep{2012AIPC.1505..277D}. Using the
model for the Galactic rotation curves by \cite{2007A&A...468..993K},
we can determine that 
the gas is located at a distance of about 11 kpc.

Assuming a ratio $X_{CO->N_{H_2}}=1.5 \times {10}^{20}$ between the CO
velocity integrated intensity and the column density of molecular gas,
$N_{H_2}$, the total column density from the extraction region of
\hessj\ is $ 1.7 \times {10}^{22}$ cm$^{-2}$. At 11 kpc, the density
and the total mass are about 100 cm$^{-3}$ and $ 2.4 \times {10}^5$
solar masses, respectively.
%


\subsection{X-Ray Observations}\label{Xray}
No candidate for an X-ray counterpart of \hessj\ was found in existing
catalogs, even when extending the search radius to 0.1$^{\circ}$ away
from the source. Two data sets from {\it Chandra} and one from
XMM-\textit{Newton} were thus inspected in order to search for an
X-ray counterpart of \hessj.

\begin{figure}
  \centering
  \includegraphics[width=5.0in]{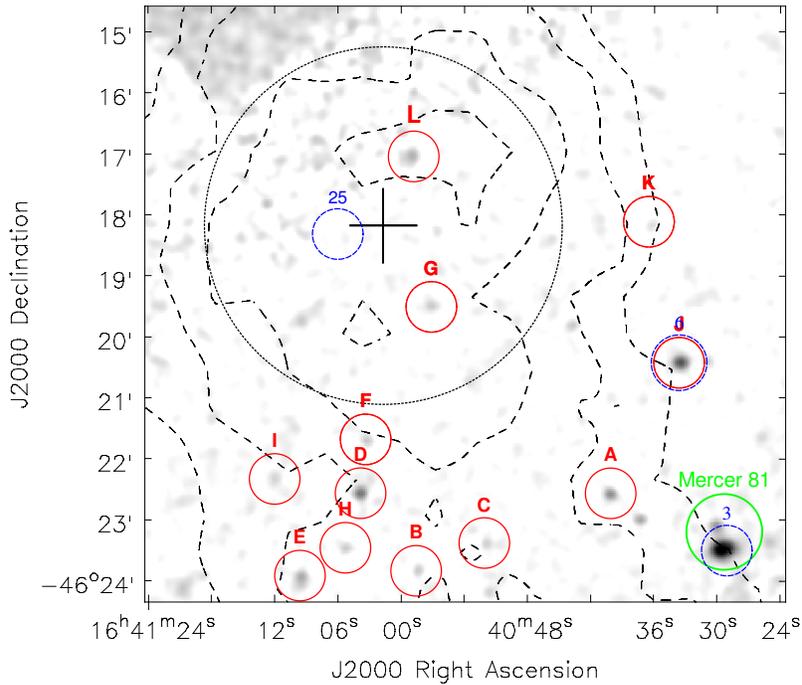}
  \caption{{\it Chandra} [1.0$-$10 keV] mosaic image of the field
    surrounding \hessj\ from the ObsIDs 11008 and 12508. The image was
    exposure-corrected, background subtracted, and smoothed with a
    Gaussian of width 10''. The best fit position of
    \hessj\ with $1\sigma$ error bars is indicated with the black
    cross, while the upper limit to the source extension
    fit is indicated by the surrounding black circle. The detected
    hard X-ray sources are shown as red circles. The blue dashed circles
    indicate the positions of the sources detected by using 
    \textit{XMM-Newton} data. The dashed contours indicate the significance of
    the VHE \g-ray emission as shown in Figure~\ref{ExcessMap}. The
    thick green circle shows the position of the stellar cluster
    Mercer 81, target of the ObsID 11008.}
  \label{Chandra_mosaic}
\end{figure}

The {\it Chandra} ObsID 11008 partially covers \hessj, with 40 ks of
exposure, while ObsID 12508 fully encloses it with 19 ks. The
data sets were processed with the CIAO package. The tool
\texttt{wavdetect} was used to identify sources, providing 32 faint
point-like or marginally extended candidates at distances
smaller than 0.1$^{\circ}$ to the \hessj\ position. This sample was
filtered by two criteria, reducing the sample to 12 candidates (see
Figure \ref{Chandra_mosaic}): first, the sources with $S/N$ ratios below 3
were rejected. Second, a cut on the hardness ratio as defined in
\cite{2009ApJS..184..158E} was applied, HR = ($H-S$)/($H+S$), where $H$
are the counts with 2$-$10 keV and $S$ the counts with 0.3$-$2 keV. The
sources with HR $\leq$ 0 were excluded. Spectral fits
using an absorbed power-law model were performed assuming a value of
$N_H$ of 2.0 $\times$ 10$^{22}$ cm$^{-2}$, corresponding to the values
reported by \cite{2005A&A...440..775K} and \cite{1990ARA&A..28..215D},
in good agreement with those derived with the NANTEN data. The
estimated flux densities in the 0.3$-$10 keV energy band result in
values from 7 $\times$ 10$^{-15}$erg cm$^{-2}$s$^{-1}$ (src. B) to 1.5
$\times$ 10$^{-13}$erg cm$^{-2}$s$^{-1}$ (src. L). No evidence of
variability was found for any of the sources after performing a
one-sample Kolmogorov-Smirnov test: the probability $P_{KS}$ for the
hypothesis of a uniform flux was $P_{KS}>0.1$. None these sources is
an obvious counterpart of \hessj\ due to their low fluxes and the lack
of any morphological feature that could point to such an association.

The \textit{XMM-Newton} ObsID 0302560201, covering the region of
\fourty\ \citep{2007ApJ...662..517F}, constitutes a partial 23.7 ks
exposure of the source area. The data set was analyzed using the XMM
SAS analysis task \texttt{edetect\_chain} simultaneously in all three
cameras and the five standard energy bands.  In this manner, 27 sources were
found, with only one consistent with the position and upper limit to
the extension of \hessj\ (see Figure~\ref{Chandra_mosaic}).
This source was detected only in the pn camera and only in the energy
band 0.5$-$1\,keV with a significance of $\approx4.6\sigma$, and it is
not detected in the {\it Chandra} data. The vignetting for this source
is 0.35 in the pn camera, so the observation is very insensitive to
the region of interest. Due to low statistics, calculating an HR or
spectrum for this source was not possible, and it is unclear whether
this may represent a counterpart.

\subsection{HE Observations}
The only High Energy (HE, 0.1$-$100 GeV) source found within
0.5$^{\circ}$ of \hessj\ is 2FGL~J1640.5$-$4633
\citep{2012ApJS..199...31N}, also present in the 10 $>$ GeV
\fermi\ Catalog as 1FHL~J1640.5$-$4634 \citep{2013ApJS..209...34A},
likely to be associated with \fourty\ \citep{2010ApJ...720..266S,
  2014ApJ...788..155G} (see Figure \ref{ExcessMap}). If the spectrum of
\hessj\ is extended to lower energies as a featureless power law, its
HE counterpart could be confused with 1FHL~J1640.5$−$4634. However,
the extrapolation of the VHE emission of \hessj\ to the \fermi\ energy
ranges predicts a flux of (5.0 $\pm$ 2.8) $\times$ 10$^{-11}$ \rm
cm$^{-2}$\,s$^{-1}$ in the 10$-$500 GeV band, a factor 10 lower than
the flux of 1FHL~J1640.5$-$4634 at those energies and thus a detection
of a GeV excess on the position of \hessj\ would imply either a
contribution from an unrelated source or from a different component of
radiation of the same source. A study to resolve such a faint,
confused source is challenging and outside the scope of this work.

\section{Discussion}\label{disc}

Possible scenarios to explain the emission from \hessj\ include the emission 
from accelerated particles within an SNR, a molecular cloud illuminated
by cosmic rays (CRs), a pulsar wind nebula (PWN) and a \g-ray
binary. These scenarios are discussed below. 

\begin{figure}
  \centering
  \includegraphics[width=5.0in]{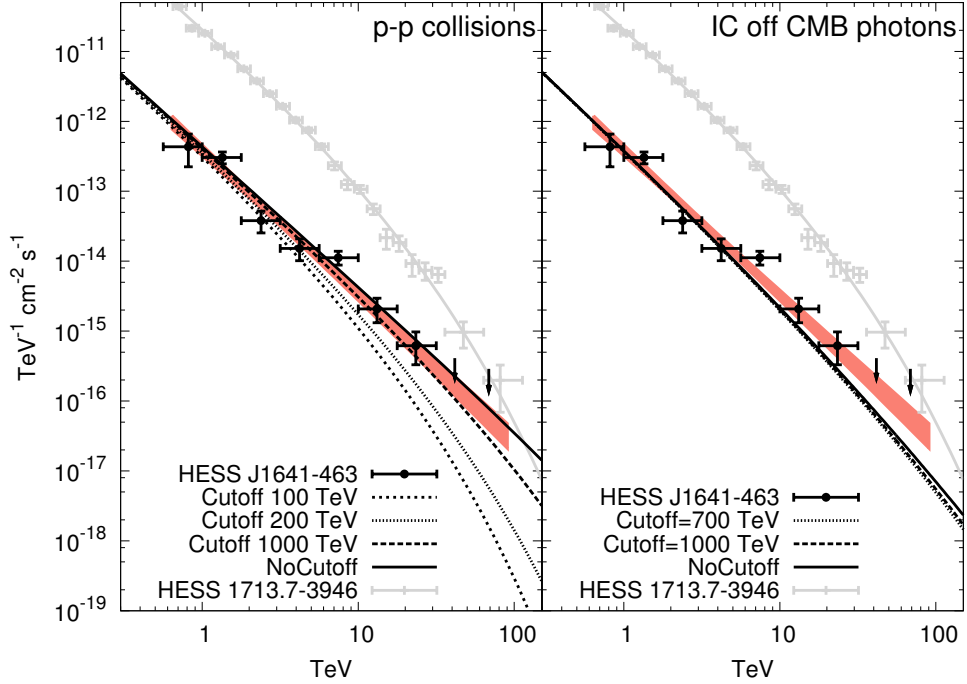}
  \caption{Differential \g-ray spectrum of \hessj\ together with
    the expected emission from p$-$p collisions (left) and IC off CMB
    photons (right). The pink area represents the $1\sigma$ confidence
    region for the fit to a power-law model, the black data points the
    H.E.S.S. measured photon flux ($1\sigma$ uncertainties), the
    arrows the 95\% CL upper limits on the flux level, and the black
    curves the expected emission from the models, assuming different
    particle energy cutoff values. For comparison, the gray data
    points and curve represent the archival spectrum and the
    corresponding best-fit model, respectively, of SNR RX
    J1713.7$-$3946 \citep{2007A&A...464..235A}.}
  \label{EmissionM}
\end{figure}

If G338.5+0.1 is a young SNR, it can accelerate particles up to
hundreds of TeV.  The left panel of Figure~\ref{EmissionM} shows the
comparison between the \hess\ spectrum and the spectrum produced by
accelerated protons from G338.5+0.1, interacting with the ambient
gas. The predicted spectra are calculated using the parameterization of
\cite{2006PhRvD..74c4018K}, assuming a proton spectrum with a
power-law slope of $-$2.1 and multiple cutoff energies. The profile of
the log-likelihood ratio test statistic \citep{2005NIMPA.551..493R}
was used to estimate a confidence interval of the cutoff energies,
while considering the spectral index and normalization as nuisance
parameters and ignoring systematic errors. The 99\% confidence level
(CL) lower limit on the cutoff energy corresponds to 100 TeV. This
proton spectrum is one of the hardest ever inferred to explain the
emission from a \g-ray source and agrees well with the prediction by
diffusive shock acceleration in young SNRs. Remarkably, the \g-ray
spectrum of \hessj\ is harder than that observed from the young SNR
RXJ1713$-$4936 at energies above few TeV, where a cutoff is seen
\citep{2007A&A...464..235A}. If the TeV luminosity measured by
\hess\ is produced by collisions of protons with the ambient gas, then
the total energy of the supernova explosion converted into hadron
acceleration is $W_p = L_{\gamma} \, t_{\pi^{0}}
  \approx{10}^{50} n^{-1}$, where $L_{\gamma}= 4 \times {10}^{34}$
  erg s$^{-1}$ is the total luminosity measured by \hess\ above 0.64 TeV (at
  11 kpc) and $t_{\pi^{0}} \approx 5 \times 10^{15} (n/1\,\mathrm{cm}^{-3})^{-1}\,
  \mathrm{s}$ is the cooling time of protons through the channel of $\pi^{0}$
  production \citep{2004vhec.book.....A}. With a proton spectrum extending almost up to 1 PeV, \hessj\ may represent a source population
contributing significantly to the galactic CR flux around the
knee.


If G338.5+0.1 is older (5$-$17 kyr; see Section \ref{radio_sec}) then VHE
protons accelerated by the young SNR\,G338.3$-$0.0, positionally
coincident with \fourty\, \citep{2014MNRAS.439.2828A}, could have
already reached the dense molecular cloud (MC) coincident with
\hessj. This would explain the relatively high brightness of
\hessj\ in comparison with \fourty\ at high energies, as shown in
Figure \ref{Slices} \citep{1996A&A...309..917A, 2009MNRAS.396.1629G}. In
such a scenario, \fourty\ would no longer look
like a {\it pevatron}, as the highest energy CRs would have already
left \citep{1996A&A...309..917A}. The much younger adjacent
G338.3$-$0.0 would be in this scenario a major source of CRs.

Electrons of hundreds of TeV IC (inverse Compton) scattering off
the cosmic microwave background photons (CMB) could explain the
emission from \hessj. These electrons would be accelerated either in
G338.5+0.1 or in the PWN associated with the young energetic
pulsar, PSR J1640$-$4631, discovered within the observational
boundaries of \fourty~\citep{2014ApJ...788..155G}. Even assuming a
pure power law for the primary electron spectrum, the cross section
for IC scattering decreases at high energies resulting in a break in
the \g-ray spectrum at multi TeV energies. Such a break is not
observed in the spectrum of \hessj. The
predicted IC radiation, shown in the right panel of
Figure \ref{EmissionM}, was obtained by assuming that the electron
cooled spectrum is a power law of spectral index $-$3.14 with
different cutoff energies. The 99\% CL lower limit on the cutoff
energy, derived as in the case of
the proton model using the exact Klein-Nishina expression for the
IC emission, corresponds to 700 TeV. It is extremely
difficult to accelerate electrons in SNRs to such energies, as hundred TeV
electrons suffer severe synchrotron losses in the amplified magnetic
fields of acceleration sites.  Both the absence of a break in the
\g-ray spectrum of \hessj\ and the derived lower limit on the cutoff
energy of the electron spectrum strongly disfavor the leptonic
scenario.

A \g-ray binary scenario could also be considered, given the
point-like morphology of \hessj\ and that a similarly hard spectral
index of $-$2.23 has been found in one of these systems (LS 5039;
\citealt{2005Sci...309..746A}). An X-ray flux as low as $\sim10^{-14}$
erg cm$^{-2}$ s$^{-1}$ is expected from a faint X-ray binary system
similar to HESS J0632+057 \citep{2009ApJ...690L.101H} assuming a
distance of 11 kpc, where the lack of an obvious optical counterpart
could be due to high optical extinction caused by the large distance and
the position close to the Galactic plane.

\section{Conclusions}
Deeper exposures with H.E.S.S. together with a study of the emission
in various energy bands made it possible to discover a new unique VHE
source, showing one of the hardest \g-ray spectra ever found at these
energies, extending up to at least 20 TeV without a break. In order to
explain the observed VHE \g-ray spectrum, scenarios where protons are
accelerated up to hundreds of TeV at either G338.5+0.1 or
G338.3$-$0.0, and then interact with local gas or nearby massive
MCs are the most compelling ones. Other possible scenarios, such as a
PWN or a \g-ray binary, are disfavored but cannot be discarded. Deeper
X-ray and VHE \g-ray observations, together with a better PSF for the
latter, would allow for a better identification of the source.

\acknowledgments 
The support of the Namibian authorities and of the University of
Namibia in facilitating the construction and operation of H.E.S.S. is
gratefully acknowledged, as is the support by the German Ministry for
Education and Research (BMBF), the Max Planck Society, the French
Ministry for Research, the CNRS-IN2P3, and the Astroparticle
Interdisciplinary Programme of the CNRS, the U.K. Science and
Technology Facilities Council (STFC), the IPNP of the Charles
University, the Polish Ministry of Science and Higher Education, the
South African Department of Science and Technology and National
Research Foundation, and by the University of Namibia. We appreciate
the excellent work of the technical support staff in Berlin, Durham,
Hamburg, Heidelberg, Palaiseau, Paris, Saclay, and in Namibia in the
construction and operation of the equipment. This research has made
use of {\it Chandra} Archival data, as well as the {\it Chandra} Source Catalog,
provided by the {\it Chandra} X-ray Center (CXC) as part of the {\it Chandra} Data
Archive. This research has made use of software provided by the
{\it Chandra} X-ray Center (CXC) in the application packages CIAO, ChIPS,
and Sherpa. This research uses on observations obtained with
XMM-\textit{Newton}, an ESA science mission with instruments and contributions
directly funded by ESA Member States and NASA. This research has made
use of the SIMBAD database, operated at CDS, Strasbourg, France.



{\it Facilities:} \facility{H.E.S.S.}.

\end{document}